\documentclass[%
 reprint,
showpacs,preprintnumbers,
 amsmath,amssymb,
 aps,
 prd,
]{revtex4-1}

\usepackage{graphicx}
\usepackage{dcolumn}
\usepackage{bm}
\usepackage{color}

\begin{document}

\title{Gravitational waves with dark matter minispikes: The combined effect  }
\author{Xiao-Jun Yue}
\email{yuexiaojun@tyut.edu.cn}
  \affiliation{College of Information Engineering, Taiyuan University of Technology, Taiyuan 030024, China\\
   and Shanghai Astronomical Observatory, Shanghai, 200030, China}
\author{Wen-Biao Han}
\email{Corresponding author: wbhan@shao.ac.cn}
\affiliation{ Shanghai Astronomical Observatory, Shanghai, 200030, China\\
and  School of Astronomy and Space Science,
University of Chinese Academy of Sciences, Beijing 100049, China}
\date{\today}

\begin{abstract}
It was shown that the dark matter(DM) minihalo around an intermediate mass black hole(IMBH) can be redistributed into a cusp, called the DM minispike. We consider an intermediate-mass-ratio inspiral consisting of  an IMBH harbored in a DM minispike with nonannihilating DM particles and a small black hole(BH) orbiting around it. We investigate gravitational waves(GWs) produced by this system and analyze the waveforms with the comprehensive consideration of gravitational pull, dynamical friction and accretion of the minispike and calculate the time difference and phase difference caused by it. We find that for a certain range of frequency, the inspiralling time of the system is dramatically reduced for smaller central IMBH and large density of DM. For the central IMBH with $10^5M_\odot$, the time of merger is ahead, which can be distinguished by LISA, Taiji and Tianqin.  We focus on the effect of accretion and compare it with that of gravitational pull and friction. We find that the accretion mass is a small quantity compared to the initial mass of the small BH and the accretion effect is inconspicuous compared with friction. However, the accumulated phase shift caused by accretion is large enough to be detected by LISA, Taiji and Tianqin, which indicate that the accretion effect can not be ignored in the detection of GWs.

\end{abstract}


\maketitle


\section{\label{s1}Introduction}
The observations of astrophysics and cosmology indicate that dark matter(DM) makes a large fraction of galaxies, but  the origin and nature are still unknown. Particle physicists seek to probe DM particles directly in laboratory, and astronomers would like to detect DM through indirect searches. The decays and annihilations of DM lead to potentially detectable fluxes of high energy radiation such as gamma rays. Several astronomical detectors, such as the Fermi Large Telescope(Fermi-LAT), the Major Atmospheric Gamma-ray Imaging Cherenkov(MAGIC) telescope have fueled a sustained interest in this domain\cite{I0}.

The distribution of DM is a subject of great interest. The one most used for cold dark matter(CDM) is the  Navarro, Frenk and White(NFW) profile\cite{I1}.   Via N-body simulations, it was  pointed out the existence of a universal density profile for DM halos. In \cite{I2}, it was shown that the density has a cusp at the center of galaxies because of the large potential well there. Gondolo and Silk present a simple Newtonian model to suggest that if a massive black hole resides at the center of the galaxy, the strong gravity could lead to a significant increase of density in the central region and create a ``spike", which enhances the DM annihilation rate\cite{I02}. The estimation of DM density in the vicinity of a massive BH in the general relativity formalism is also proposed in \cite{I3}. Other studies show that some events  such as merges of host galaxies can make the DM spike weakened\cite{I4,I5,I6,I7}, which make this issue controversial. On the other hand, the intermediate mass black hole(IMBH) with a mass range between $10^2 M_\odot$ and $10^6 M_\odot$ may have a DM minispike as it is less likely to experience mergers in the past\cite{I8,I9}, which may be an ideal place for DM detection. Especially, the spin of IMBH can actually enhance the spike \cite{Ferrer17}. Other DM models have different paradigms of the nature of DM particles from CDM, as has been proposed and explored widely for different candidates such as self-interacting dark matter \cite{I91},
warm dark matter \cite{I92,I93}, axion/scalar, or wave dark matter \cite{I94,I95}.

The discovery of gravitational waves(GW) by the ground based detectors such as LIGO and VIRGO\cite{I96} has opened a new observational window on the detection of Universe. The future space-based detectors, such as LISA, Taiji and Tianqin\cite{I97} will surely facilitate us achieving more observational programs. Whether the DM mass distribution could have an influence on the orbits of stars and other objects, such as BHs and neutron stars, which can leave a sign in the gravitational wave is an important issue. In \cite{I100} it was pointed out that the DM minispike could impact the GW waveform, which can be detected by LISA. But in \cite{I11} the authors gave a wide survey of the environmental corrections such as as electric charges, magnetic fields, accretion disks and dark matter halos to the GW signals with the order of magnitude estimates and conclude that environmental effects are typically negligible for most LISA sources.  However, the subsequent study using filtering technique and Fisher matrix analysis  indicates that  the environmental effects do affect GW detectability and the DM parameters can be measured by LISA quite accurately\cite{I12}. In\cite{I10} a particular system of a stellar object as a test particle inspiralling into some compact configuration of DM clouds was studied.   Recently, the possible impact of DM on the GW signals from neutron star mergers is also studied in\cite{I13}.

The gamma ray observation of DM rely on the weakly interaction of DM particles, while the GW detection can be applicable for noninteracting DM. Previous studies focus on different aspects of the DM effect to GW. Reference\cite{I10} reveals the potential importance of the dynamical friction and accretion of DM configuration on the GW waveform. In \cite{I100} the effect of gravitational force of the DM minispike around a central IMBH was studied in detail and \cite{I12} indicated the significant difference of GW signal made by  friction of the minispike.  On the other hand, with the DM minispike the accretion is inevitable, but whether and what extend the accretion effect can influence the GW waveform is still a question.  In this paper we investigate the combined effect of gravitational pull, dynamical friction and accretion of DM minispike.  We consider an intermediate-mass-ratio inspirals in the DM minispike and calculate the GW waveform. We concentrate on accretion, find out its effect and compare it with other effects.

This paper is organized as follows. In Sec.\ref{s2} we derive the dynamical equations. In Sec.\ref{s3} the GW waveform is calculated analytically and numerically and we conclude in Sec.\ref{s4}.

\section{\label{s2}Dynamical Equations}
In this section, we review the minispike model concisely and then derive the dynamical equations.
 We employ the same minispike model as in\cite{I12} and a more detailed description can be found there. The initial DM density is assumed to be $\rho\propto r^{-\alpha_{\rm ini}}$. After the adiabatic growth of the IMBH the DM profile is described by\cite{I02}
 \begin{equation}
 \rho_{\rm DM}(r)=\rho_{\rm sp}\left(\frac{r_{\rm sp}}{r}\right)^{\alpha},(r_{\rm min}\leq r\leq r_{\rm sp}),\label{e1}
 \end{equation}
 where $r_{\rm min}$ is the minimum of the stable circular orbit  of the central IMBH $r_{\rm min}=r_{ISCO}=6\pi G M_{\rm BH}/c^2$ and $M_{\rm BH}$ is the mass of the IMBH. $r_{\rm sp}$ is defined by $r_{\rm sp}\sim 0.2 r_h$ and $r_h$ is the influence of the central IMBH, which is defined by $4\pi\int_0^{r_h}\rho_{\rm DM}r^2 dr=2M_{\rm BH}$. $\rho_{\rm sp}$ is the normalization constant which is the DM density in $r_{\rm sp}$. $\alpha$ is the slope of the minispike, $\alpha=(9-2\alpha_{\rm ini})/(4-\alpha_{\rm ini})$. Beyond the spike radius $r_{\rm sp}$, we assume the
DM distribution is the NFW profile
\begin{equation}
\rho_{\rm NFW}=\frac{\rho_s}{(r/r_s)(1+r/r_s)},\label{e2}
\end{equation}
 where $\rho_s$ and $r_s$ are parameters related to cluster mass and concentration parameters \cite{I1}.

 For an IMBH with mass of  $M_{\rm BH}=10^3 M_\odot$ and the total mass of the DM minihalo is $M_{halo}=10^6 M_\odot$, the parameters $\rho_{\rm sp}$ and $r_{\rm sp}$ are given to be $\rho_{\rm sp}=226M_\odot/pc^3$ and $r_{\rm sp}=0.54pc$. The slope of the DM minispike has different values in different cases. As in \cite{I12,I100}, here we assume $\alpha$ to vary between $1.5\leq\alpha\leq3$.

 With the model of DM minispike, we can now derive the dynamical equations of the binary system. Here we consider the IMBH in the center of DM minihalo with the mass of $1000M_{\odot}$ and a small black hole with the mass $\mu=10M_{\odot}$ orbiting around it. The total mass $M=M_{\rm BH}+m\approx10^3M_{\odot}=M_{\rm BH}$ and the reduced mass $m=\frac{M_{\rm BH} m}{M_{\rm BH}+m}\approx M_{\odot}= m$. The barycenter is approximately the mass center of the IMBH. When we consider the relative motion of the two objects, we have to add the gravitational pull of the minihalo around the IMBH. So the equation of motion in the radial direction is
 \begin{equation}
\dot{\mu}\dot{r}+\mu\ddot{r}-\mu r\dot{\theta}^2=-\frac{G\mu M_{\rm eff}}{r^2}-\frac{\mu F}{r^{\alpha-1}},\label{e3}
\end{equation}
where
\begin{eqnarray}
M_{\rm eff}&&=
\begin{cases}
M_{\rm BH}-M_{\rm DM}(<r_{\rm min}),& \text{$r_{\rm min}<r<r_{\rm sp}$},\\
M_{\rm BH},& \text{$r<r_{\rm min}$},\\
\end{cases}\\\label{e4}
F&&=
\begin{cases}
G r_{\rm min}^{\alpha-3}M_{\rm DM}(<r_{\rm min}),& \text{$r_{\rm min}<r<r_{\rm sp}$},\\
0,& \text{$r<r_{\rm min}$}
\end{cases}\label{e5}
\end{eqnarray}
where $M_{\rm DM}(<r_{\rm min})=4\pi r_{\rm sp}^\alpha\rho_{\rm sp}r_{\rm min}^{\alpha-3}/(3-\alpha)$ is the DM contained in $r_{ISCO}$. The first term on the right is the effective mass of IMBH corrected by DM. The second is the gravitational effect of DM.

For the circular orbit, $\dot{r},\ddot{r}=0$, so we have
\begin{equation}
\dot{\theta}=\omega_s=\sqrt{\frac{GM_{\rm eff}}{r^3}+\frac{F}{r^\alpha}},\label{e6}
\end{equation}
which is the same as the Kepler's law but modified by the DM minispike. Of course the orbit cannot keep a circular shape as the GW radiation and dissipation of friction and accretion. In the following we will derive the time evolution of the orbital radius and the orbit can be regarded as a quasicircular orbit.

Taking account of the  GW back reaction, the dynamical friction and the variation of the mass of the small BH led by accretion, the equation of motion in the tangential direction turns out to be
\begin{equation}
2\mu\dot{r}\dot{\theta}+\mu r\ddot{\theta}=-F_{\rm GW}-F_{\rm DF}-\dot{\mu}r\dot{\theta},\label{e7}
\end{equation}
where $F_{\rm GW}$ is the force acted by the gravitational wave and $F_{\rm DF}$ is the force of dynamical friction. The third term of the right $\dot{\mu}r\dot{\theta}$ is the drag force due to accretion. In the quasicircular orbit condition and the intermediate-mass-ratio inspirals,
\begin{equation}
F_{\rm GW}=\frac{1}{\omega_s r}\frac{dE_{\rm GW}}{dt}=\frac{1}{\omega_s r}\frac{32}{5}\frac{G\mu^2}{c^5}r^4\omega_s^6,\label{e8}
\end{equation}
where $dE_{\rm GW}/dt=\frac{32}{5}\frac{G\mu^2}{c^5}r^4\omega_s^6$ is the gravitation radiation power in the quadruple formula. The dynamical friction $F_{\rm DF}$ is sometimes called the gravitational drag developed by Chandrasekhar\cite{II2}. When the small BH moves through the DM minispike, the gravitational field generated by the DM is felt universally, being tantamount to a net decelerating force acting on it. The dynamical friction force is given by\cite{II3}
\begin{equation}
F_{\rm DF}=\frac{4\pi G^2\mu^2\rho_{\rm DM}(r)\ln\Lambda}{v^2}=\frac{4\pi G^2\mu^2\rho_{\rm DM}(r)\ln\Lambda}{r^2\omega_s^2},\label{e9}
\end{equation}
where $v$ is the velocity of the small BH and $\ln\Lambda$ is  related to  the maximum impact parameter and the typical velocity of the small BH. Here we choose $\ln\Lambda=3$, the same as \cite{I12}.

Generally speaking, when we take account of accretion, we have to distinguish whether the radius of the small BH is larger or smaller than the mean free path of the DM particles. The former is called the Bondi case and the latter is the collisionless case\cite{I10,II4}. In the Bondi case, the DM compressibility have to be taken into account which relates to the speed of sound. However, in this paper we consider the nonannihilating DM particles and the interactions except gravitation is out of our consideration. So the accretion is described by
\begin{equation}
\dot{\mu}=\sigma\rho_{\rm DM}v,\label{e10}
\end{equation}
where $\sigma$ is the accretion cross section.If the small stellar object is a black hole and the DM are treated as point particles, the accretion cross section is given by\cite{II5}
\begin{eqnarray}
&&\sigma=\frac{\pi G^2\mu^2}{c^4\cdot(v^2/c^2)}\nonumber\\
&&\times\left\{ \frac{[8(1-v^2/c^2)]^3}{4(1-4v^2/c^2+(1+8v^2/c^2)^{1/2})(3-(1+8v^2/c^2)^{1/2})^2}\right\}.\nonumber\\
\label{e11}
\end{eqnarray}
The quantity in curly brackets is a slowly varying function of $v$ going from $16$ for $v=0$ to $27$ for $v=c$.In the regime of our consideration, $v\ll c$ and we can expand the term in the curly brackets to be one order in $v^2/c^2$:$\sigma=(16\pi G^2 \mu^2)(1+v^2/c^2)/(v^2c^2)$, then we have
\begin{equation}
\dot{\mu}=\frac{16\pi G^2\mu^2\rho_{\rm DM}}{c^2v}(1+\frac{v^2}{c^2}).\label{e12}
\end{equation}
The ratio of the accretion drag force with the dynamical friction is $\dot{\mu}v/F_{\rm DF}= 4v^2(1+v^2/c^2)/(c^2\ln\Lambda)\sim v^2/c^2$, so the drag of accretion is a small quantity  compared to friction when $v\ll c$.
With Eqs.(\ref{e8},\ref{e9},\ref{e12}), after some algebra and simplification, Eq.(\ref{e7}) can be rewritten to be
\begin{eqnarray}
&&2\mu r\omega_s^2\dot{r}+\mu r^2\omega_s\dot{\omega_s}\nonumber\\
=&&-\frac{32}{5}\frac{G\mu^2}{c^5}r^4\omega_s^6-\frac{4\pi G^2\mu^2\rho_{\rm sp}r_{\rm sp}^\alpha \ln\Lambda}{r^{\alpha+1}\omega_s}(1+\frac{\dot{\mu}v}{F_{\rm DF}})\nonumber\\
=&&-\frac{32}{5}\frac{G\mu^2}{c^5}r^4\omega_s^6-\frac{4\pi G^2\mu^2\rho_{\rm sp}r_{\rm sp}^\alpha \ln\Lambda}{r^{\alpha+1}\omega_s}(1+b_A)
\label{e13}
\end{eqnarray}
where the function
\begin{equation}
b_A=\frac{4r^2\omega_s^2}{c^2 \ln\Lambda}(1+\frac{r^2\omega_s^2}{c^2})\label{e14}
\end{equation}
is the ratio of the accretion drag with dynamical friction. For convenience, as in \cite{I12} we introduce the dimensionless radius parameter $x$ defined by
\begin{eqnarray}
x=&&\epsilon^{1/(3-\alpha)}r,\label{e15}\\
\epsilon=&&\frac{F}{GM_{\rm eff}},\label{e16}
\end{eqnarray}
and one can verify that $\epsilon\ll 1$. With this Eq.(\ref{e13}) can be rewritten to be
\begin{eqnarray}
&&\frac{dx}{dt}=-c_{\rm GW}\frac{(1+x^{3-\alpha})^3}{4x^3[1+(4-\alpha)x^{3-\alpha}]}\nonumber\\
&&-c_{\rm DF}\frac{1}{(1+x^{3-\alpha})^{1/2}[1+(4-\alpha)x^{3-\alpha}]x^{-5/2+\alpha}}(1+b_A)\nonumber\\
\label{e17}
\end{eqnarray}
where the coefficients are defined by
\begin{eqnarray}
&&c_{\rm GW}=\frac{256}{5}\left(\frac{G\mu}{c^3}\right) \left(\frac{GM_{\rm eff}}{c}\right)^2 \epsilon^{4/(3-\alpha)},\label{e18}\\
&&c_{\rm DF}=(8\pi G^2\mu\rho_{\rm sp}r_{\rm sp}^\alpha \ln\Lambda)(GM_{\rm eff})^{-3/2}\epsilon^{(2\alpha-3)/[2(3-\alpha)]}.\nonumber\\
\label{e19}
\end{eqnarray}
Here the form of the function $b_A$ is kept invariant as Eq.(\ref{e14}) and we do not write it in the variable $x$. In the following we will calculate the GW waveform  and it is suitable to convert its form into the frequency domain directly in the next section. The accretion effect is manifested by the term $b_A$ as well as the coefficients of Eqs.(\ref{e18},\ref{e19}), where the mass $\mu$ varies with time. For convenience Eq.(\ref{e17})can be written as
\begin{eqnarray}
&&\frac{dx}{dt}=-c_{\rm GW}f_{\rm GW}(x)-c_{\rm DF}f_{\rm DF}(x)(1+b_A)\nonumber\\
&&=-c_{\rm GW}f_{\rm GW}(x)\left[1+\frac{c_{\rm DF}}{c_{\rm GW}}\frac{f_{\rm DF}}{f_{\rm GW}}(1+b_A)\right]\nonumber\\
&&=-c_{\rm GW}f_{\rm GW}(x)\left[1+\tilde{c}J(x)(1+b_A)\right],\label{e20}
\end{eqnarray}
where
\begin{eqnarray}
\tilde{c}&&=\frac{c_{\rm DF}}{c_{\rm GW}}\label{e21}\\
J(x)&&=\frac{4x^{11/2-\alpha}}{(1+x^{3-\alpha})^{7/2}}.\label{e22}
\end{eqnarray}

\section{\label{s3}GW waveform}

In this section we will calculate the GW waveform with the stationary phase method. In the quadrupole approximation, there are two polarizations $h_+$ and $h_\times$ in the GW waveform. For simplicity, we consider a GW coming from the optimal direction for $+$ mode. So the GW waveform is
\begin{equation}
h(t)=h_+(t)=A(t_{ret})cos[\Phi(t_{ret})],\label{e23}
\end{equation}
where $t_{ret}=t-D/c$ is the retarded time and $D$ is the distance to the source, $A(t)$ is the time dependent amplitude and $\Phi(t)$ is the time dependent GW phase, which has the form
\begin{eqnarray}
&&A(t)=\frac{1}{D}\frac{4G\mu(t)\omega_s(t)^2 r(t)^2}{c^4}\frac{1+cos^2i}{2},\label{e24}\\
&&\Phi(t)=\int^t \omega_{\rm GW}(t')dt'.\label{e25}
\end{eqnarray}
The $r(t)$ is the orbital radius, $i$ is the inclination angle which is the angle between the line of sight and the axis of the orbit. $\omega_{\rm GW}$ is the GW frequency and has the relation $\omega_{\rm GW}=2\omega_s$. For convenience we would like to work directly with the Fourier transformation of $h(t)$,
\begin{equation}
\tilde{h}(f)=\int^\infty_{-\infty} h(t)e^{2\pi ift}dt.\label{e25}
\end{equation}
This can be computed using the stationary phase method\cite{III1}. Given the function Eq.(\ref{e23}), where $d \ln A/dt\ll d\Phi(t)/dt$ and $d^2\Phi/dt^2\ll(d\Phi/dt)^2$, the stationary phase approximation provide the following estimation of the Fourier transformation:
\begin{eqnarray}
&&\tilde{h}(f)\approx\frac{1}{2}e^{i\Psi(t)}A(t)\left[\frac{df}{dt}\right]^{-1/2}\label{e26}\\
&&\Psi(t)=2\pi f\frac{D}{c}+\tilde{\Phi}(t)-\frac{\pi}{4}\label{e27}\\
&&\tilde{\Phi}(t)=2\pi ft-\Phi(t)\label{e28}
\nonumber
\end{eqnarray}
We have to work in the explicit frequency domain while the above expressions have the variable $t$. Now we employ the same method as in \cite{I12} to rewrite them in the frequency domain.
\subsection{\label{s41}GW waveform in the frequency domain}

The time $t$ is related to frequency by $2\pi f=\omega_{\rm GW}(t)$ so we have
\begin{eqnarray}
&&\frac{df}{dt}=\frac{1}{2\pi}\frac{d\omega_{\rm GW}}{dt}=\frac{1}{\pi}\frac{d\omega_s}{dt}\nonumber\\
&&=-\frac{1}{\pi}(GM_{\rm eff})^{1/2}\epsilon^{3/[2(3-\alpha)]}\frac{3+\alpha x^{3-\alpha}}{x^{5/2}(1+x^{3-\alpha})^{1/2}}\frac{dx}{dt}\nonumber\\
&&=\frac{1}{\pi}(GM_{\rm eff})^{1/2}\epsilon^{3/[2(3-\alpha)]}c_{\rm GW}\nonumber\\
&&\times\left[1+\tilde{c}J(x)\left(1+b_A\right)\right]\frac{3}{4}x^{-11/2}K(x),\nonumber\\
\label{e29}
\end{eqnarray}
where the Eq.(\ref{e6}) and Eq.(\ref{e20}) was used. Except the $\mu(t)$ in the coefficient $c_{\rm GW}$ Eq.(\ref{e18})is a function of time $t$, the upper expression is a function of $x$ but not $f$. In the following we have to transform the independent variable $x$ into $f$.
\begin{eqnarray}
f=&&\frac{\omega_{\rm GW}}{2\pi}=\frac{1}{\pi}\left[\frac{GM_{\rm eff}}{r^{3}}+\frac{F}{r^\alpha}\right]^{1/2}\nonumber\\
=&&\frac{\sqrt{GM_{\rm eff}}}{\pi}r^{-3/2}\left[1+\frac{1}{2}r^{3-\alpha}\epsilon-\frac{1}{8}r^{2(3-\alpha)}\epsilon^2+\cdot\cdot\cdot\right].\nonumber\\
\label{e30}
\end{eqnarray}
In the last step the frequency $f$ is expanded in a Taylor series in power of $r$. Inverting this equation we can expand $r$ in $\epsilon$:
\begin{eqnarray}
r=\delta^{1/(3-\alpha)}\left[1+\frac{1}{3}\delta\epsilon+\frac{2-\alpha}{9}\delta^2\epsilon^2+\cdot\cdot\cdot\right]\label{e31}\\
\delta=\left(\frac{GM_{\rm eff}}{\pi^2f^2}\right)^{(3-\alpha)/3}.\label{eq32}
\end{eqnarray}
Introducing the new variable
\begin{equation}
\chi=1+\frac{1}{3}\delta\epsilon+\frac{2-\alpha}{9}\delta^2\epsilon^2+\cdot\cdot\cdot,\label{e34}
\end{equation}
we can get from the definition of $x$ Eq(\ref{e15})that
\begin{eqnarray}
x=(\delta\epsilon)^{1/(3-\alpha)}\chi\label{e33}
\end{eqnarray}
With the upper two equations we can transform any function of $x$ into a function of $f$. Then
\begin{eqnarray}
&&\frac{df}{dt}\nonumber\\
&&=\frac{96}{5}(GM_{\rm eff})^{2/3}G^{5/3}\mu(t)c^{-5}\pi^{8/3}f^{11/3}\chi^{-11/2}\nonumber\\
&&\times\left[K(1+\tilde{c}J\left(1+\tilde{b}_A\right)\right]\nonumber\\
&&=\frac{3}{5}\pi\left(\frac{8\pi GM_c(t)}{c^3}\right)^{5/3}f^{11/3}\chi^{-11/2}\left[K(1+\tilde{c}J\left(1+\tilde{b}_A\right)\right],\nonumber\\
\label{e35}
\end{eqnarray}where in the first step we used Eq.(\ref{e18}) and the second step we used the definition of chirp mass $M_c=M_{\rm eff}^{2/5}\mu^{3/5}$. The $\tilde{b}_A$ is the function $b_A$ rewritten in the new variable $f$ and $\chi$
\begin{equation}
\tilde{b}_A=\frac{4(GM_{\rm eff})^{2/3}(\pi^2f^2)^{1/3}\chi^2}{ c^2 \ln\Lambda}\left(1+\frac{(GM_{\rm eff})^{2/3}(\pi^2f^2)^{1/3}\chi^2}{ c^2}\right).\label{e36}
\end{equation}

Next we rewrite the mass of the small BH $\mu(t)$ in the frequency domain. From Eq.(\ref{e12}) we have

\begin{eqnarray}
d\mu=&&\frac{16\pi G^2\mu^2\rho_{\rm DM}}{c^2v}(1+\frac{v^2}{c^2})dt=\frac{16\pi G^2\mu^2\rho_{\rm DM}}{c^2r\omega_s}(1+\frac{r^2\omega_s^2}{c^2})dt\nonumber\\
=&&\frac{16\pi G^2\mu^2\rho_{\rm sp}r_{\rm sp}^{\alpha}}{c^2r^{\alpha+1}\pi f}\left(1+\frac{(GM_{\rm eff})^{2/3}(\pi^2f^2)^{1/3}\chi^2}{ c^2}\right)\frac{dt}{df}df\nonumber\\
=&&\frac{16\pi G^2\mu^2\rho_{\rm sp}r_{\rm sp}^\alpha(1+(GM_{\rm eff})^{2/3}(\pi^2f^2)^{1/3}\chi^2/ c^2)}{c^2\chi^{\alpha+1}(GM_{\rm eff}/(\pi^2 f^2))^{(\alpha+1)/3}\pi f}\nonumber\\
&&\times\frac{5}{3\pi}(8\pi GM_{\rm eff}^{2/5})^{-5/3}\mu^{-1}\frac{\chi^{11/2}}{f^{11/3}[K(1+\tilde{c}_1J_1(1+\tilde{b}_A))]}df.\nonumber\\
\label{e37}
\end{eqnarray}
In the last step we used Eq.(\ref{e35}).  The upper equation has the solution
\begin{eqnarray}
&&\mu(f')\nonumber\\
=&&\mu_0exp\left[\int_{f}^{f'}\frac{16\pi G^2\rho_{\rm sp}r_{\rm sp}^\alpha(1+(GM_{\rm eff})^{2/3}(\pi^2f''^2)^{1/3}\chi^2/ c^2)}{c^2\chi^{\alpha+1}(GM_{\rm eff}/(\pi^2 f''^2))^{(\alpha+1)/3}\pi f''}\right.\nonumber\\
&&\left.\times\frac{5}{3\pi}(8\pi GM_{\rm eff}^{2/5})^{-5/3}\frac{\chi^{11/2}}{f''^{11/3}[K(1+\tilde{c}_1J_1(1+\tilde{b}_A))]}df''\right],\nonumber\\
\label{e38}
\end{eqnarray}
where $\mu_0$ is the initial mass of the small BH with GW frequency $f$ and $\mu(f')$ is the mass with frequency $f'$. Using Eq.(\ref{e35}) and Eq.(\ref{e38}), we get
\begin{eqnarray}
\tilde{\Phi}(f)&&=\frac{10}{3}\left(\frac{8\pi GM_{\rm eff}^{2/5}}{c^3}\right)^{-5/3}\nonumber\\
&&\times\left[-f\int_{f_c}^f df'\frac{\chi^{11/2}}{\mu(f')f'^{11/3}K(1+\tilde{c}J(1+\tilde{b}_A))}\right.\nonumber\\
&&\left.+\int_{f_c}^f df'\frac{\chi^{11/2}}{\mu(f')f'^{8/3}K(1+\tilde{c}J(1+\tilde{b}_A))}.\right]\nonumber\\
\label{e39}
\end{eqnarray}
The integral limit of the integration $f_c$ is the upper bound of the LISA frequency band, and $f_c>f$. At last, we substitute Eqs.(\ref{e6},\ref{e15},\ref{e35}) to Eqs.(\ref{e26},\ref{e27},\ref{e28}) we can get the final form
\begin{eqnarray}
&&\tilde{h}(f)=\mathcal{A}f^{-7/6}e^{i\Psi(f)}\chi^{19/4}\left[ K(x)\left(1+\tilde{c}_1J(x)(1+\tilde{b}_A)\right)\right]^{-1/2}\nonumber\\
\label{e391}\\
&&\mathcal{A}=\left(\frac{5}{24}\right)^{1/2}\frac{1}{\pi^{2/3}}\frac{c}{D}\left(\frac{GM_c(f)}{c^3}\right)^{5/6}\frac{1+cos^2i}{2}\nonumber\\
\label{e40}\\
&&\Psi(f)=2\pi f\left(t_c+\frac{D}{c}\right)-\Phi_c-\frac{\pi}{4}-\tilde{\Phi}(f) 
\label{e41}
\end{eqnarray}
where $t_c$ and $\Phi_c$ is the time and phase at $f_c$ and the $\tilde{\Phi}(f)$ has the form of Eq.(\ref{e39}). Note that when a DM minispike is not present around the IMBH, $\chi\rightarrow 1$, $K\rightarrow 1$, $M_c\rightarrow M_{c0}=\mu_0^{3/5}M_{\rm BH}^{2/5}$ and the waveform becomes
\begin{eqnarray}
&&\tilde{h}(f)=\mathcal{A}f^{-7/6}e^{i\Psi(f)}
\label{e42}\\
&&\mathcal{A}=\left(\frac{5}{24}\right)^{1/2}\frac{1}{\pi^{2/3}}\frac{c}{D}\left(\frac{M_{c0}}{c^3}\right)^{5/6}\frac{1+cos^2i}{2}\nonumber\\
\label{e43}\\
&&\Psi(f)=2\pi f\left(t_c+\frac{D}{c}\right)-\Phi_c-\frac{\pi}{4}-\tilde{\Phi}_0(f)\nonumber\\
\label{e44}\\
&&\tilde{\Phi}_0(f)\nonumber\\
&&=\left(8\pi\frac{GM_{c0}}{c^3} \right)^{-5/3}\left(-\frac{3}{4}f^{-5/3}-\frac{5}{4}f\cdot f_c^{-8/3}+2f_c^{-5/3}\right)\nonumber\\
\label{e45}
\end{eqnarray}

\subsection{\label{s3B}$\delta\epsilon$ expansion}
In the former subsection, we have rewritten the GW waveform in the new variable $\chi$ and frequency $f$, and from Eqs.(\ref{e16},\ref{e34},\ref{e33}) we can see that the $\delta\epsilon$ is actually the deviation from the case without DM minispike. In the range of $10^{-3}\leq f\leq10^{-1}$ which is in the LISA's detection band, with the parameters of $\rho_{\rm sp}=226M_\odot/pc^3$ and $r_{\rm sp}=0.54pc$ one can verify that $\delta\epsilon\ll1$, so the $\delta\epsilon$ can be treated as a small quantity and we can expand the GW waveform in it. First, we expand the mass of the small BH $\mu(f)$ up to the first order in $\delta\epsilon$ using Eq.(\ref{e34}):
\begin{eqnarray}
&&\mu(f')\nonumber\\
=&&\mu_0exp\left[\int_{f}^{f'}\frac{16\pi G^2\rho_{\rm sp}r_{\rm sp}^\alpha(1+(GM_{\rm eff})^{2/3}(\pi^2f''^2)^{1/3}/ c^2)}{c^2(\frac{GM_{\rm eff}}{\pi^2 f''^2})^{(\alpha+1)/3}\pi f''}\right.\nonumber\\
&&\left.\times \frac{5}{3\pi}(8\pi GM_{\rm eff}^{2/5})^{-5/3}f''^{-11/3}L^{-1}(f'')df''\right],\nonumber\\
\label{e46}
\end{eqnarray}
 where
 \begin{eqnarray}
&&L(f)=1+4c_\epsilon\tilde{\delta}^{(11-2\alpha)/[2(3-\alpha)]}(1+b_\epsilon),\label{e47}\\
&&\tilde{\delta}=\left(\frac{G}{\pi^2f^2}\right)^{(3-\alpha)/3}\label{e48}\\
&&c_\epsilon=\frac{5\pi}{32} c^5G^{-5/2}M_{\rm eff}^{-(\alpha+5)/3}\rho_{\rm sp}r_{\rm sp}^\alpha \ln\Lambda\label{e49}\\
&&b_\epsilon=\frac{(\pi^2f^2)^{1/3}(GM_{\rm eff})^{2/3}}{c^2\ln\Lambda}\left(1+\frac{(GM_{\rm eff})^{2/3}(\pi^2f^2)^{1/3}}{ c^2}\right)\nonumber\\
\label{e50}
\end{eqnarray}

 As the high precision of the sensitivity of eLISA to the GW phase, we expand $\tilde{\Phi}$ up to the first order in $\delta\epsilon$ ,
\begin{eqnarray}
&&\tilde{\Phi}=\frac{10}{3}\left(\frac{8\pi GM_{\rm eff}^{2/5}}{c^3}\right)^{-5/3}\nonumber\\
&&\times\left[-f\int_{f_{c}}^f df'\mu(f')^{-1}f'^{-11/3}L^{-1}(f')\right.\nonumber\\
&&\left.+\int_{f_{c}}^f df'\mu(f')^{-1}f'^{-8/3/3}L^{-1}(f')\right],\label{e51}
\end{eqnarray}
where for the $\mu(f')$ we use Eq.(\ref{e46}).

\subsection{\label{s3C}Time difference and phase difference}
\begin{figure}[!h]
\centering
\includegraphics[width=9.5cm]{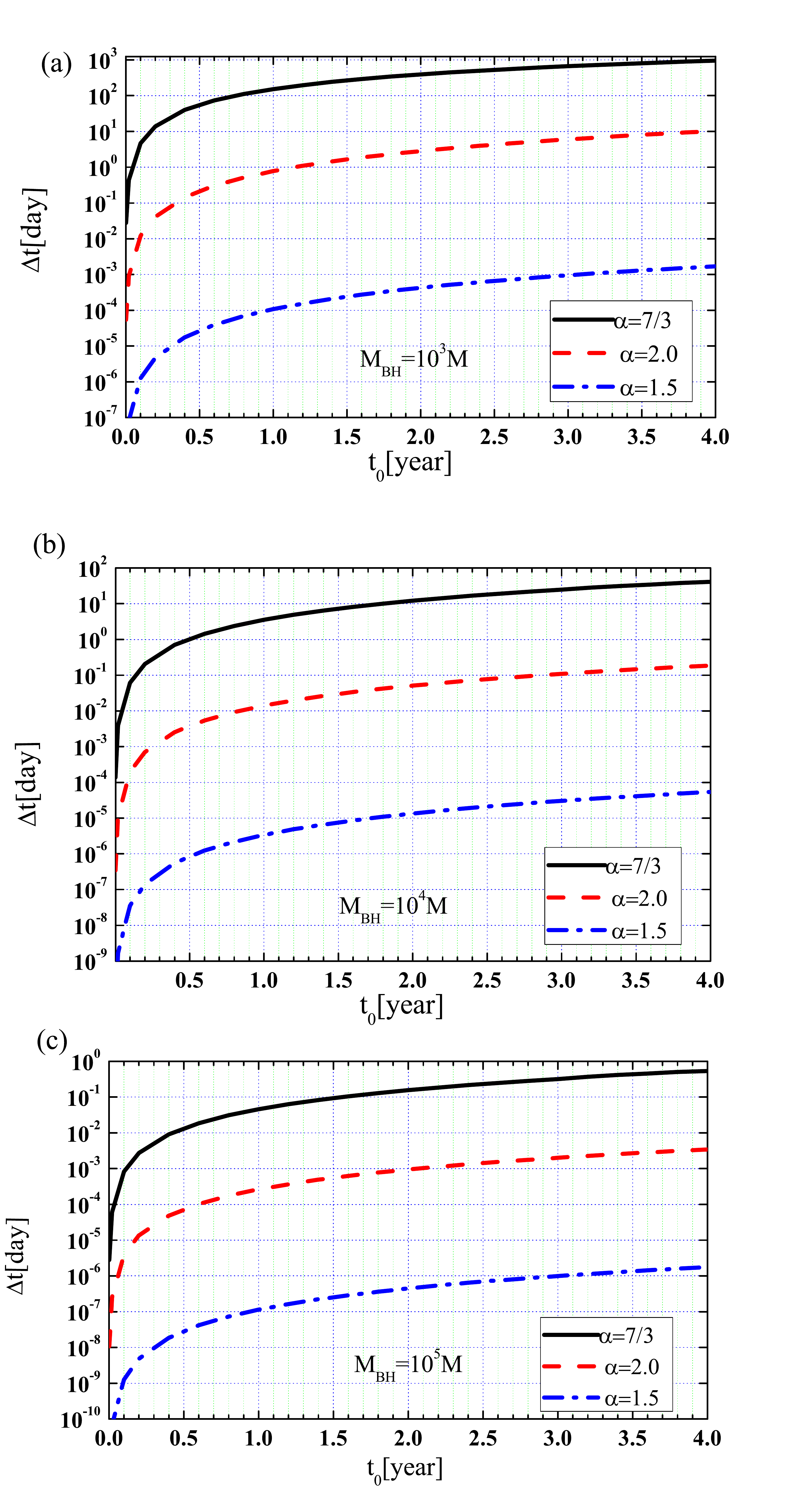}
\caption{The time difference with different central IMBH. The horizontal axis is the time without DM minispike $t_0$, the vertical axis is the time difference $\Delta t$. We take $\mu_0=10M_{\odot}$, $\rho_{\rm sp}=226M_\odot/pc^3$, $r_{\rm sp}=0.54pc$. The solid line $\alpha=7/3$, the dashed line $\alpha=2.0$, the dash-dot line $\alpha=1.5$. In figure(a) $M_{\rm BH}=10^3M_{\odot}$ and $f_c=0.1Hz$, in figure(b) $M_{\rm BH}=10^4M_{\odot}$ and $f_c=0.1Hz$, in figure(c) $M_{\rm BH}=10^5M_{\odot}$ and $f_c=f_{ISCO}$.} \label{fig01}
\end{figure}
In last subsection we derived the phase with DM minispike in the frequency space. In this subsection we provide a visualized description of the effect of DM minispike. From Eq.(\ref{e17}) we can see that the friction and accretion effect increase the velocity of the small BH falling into the central IMBH. In other words, with a certain range of frequency, the orbiting period of the small BH is reduced by the DM minispike. From Eq.(\ref{e35}) and using the $\delta\epsilon$ expansion we have

\begin{eqnarray}
t=\frac{5}{3\pi}\left(\frac{8\pi GM_{\rm eff}^{2/5}}{c^3}\right)^{-5/3}\int_{f}^{f_c} df'\mu(f')^{-1}f'^{-11/3}L^{-1}(f'),\nonumber\\
\label{e52}
\end{eqnarray}
which is the time from the frequency $f$ to $f_c$ and $f<f_c$. Without DM minispike $\mu(f')\rightarrow\mu_0$, $L(f')\rightarrow 1$ we have
\begin{eqnarray}
t_0&&=\frac{5}{3\pi}\left(\frac{8\pi G M_{c0}}{c^3}\right)^{-5/3}\int_{f}^{f_c} df'f'^{-11/3}\nonumber\\
&&=-\frac{5}{8\pi}\left(\frac{8\pi G M_{c0}}{c^3}\right)^{-5/3}(-f^{-8/3}+f_c^{-8/3}).\label{e53}
\end{eqnarray}
From Eq.(\ref{e53}) we can get $f$ as a function of $t_0$ and substitute into Eq.(\ref{e52}) we can obtain $t$ as a function of $t_0$. We define
\begin{equation}
\Delta t=t_0-t\label{e54}
\end{equation}
as the time difference of the period from $f$ to $f_c$ with and without DM minispike.

Figure \ref{fig01} depicts the time difference $\Delta t$ varies with $t_0$ with different central IMBH. Taking into account the frequency band of LISA, we set the final frequency $f_c=0.1Hz$. In Fig.\ref{fig01}(c) with the mass of the central IMBH $M_{\rm BH}=10^5 M_\odot$, the frequency of the innermost stable circular orbit $f_{ISCO}\sim 0.043<0.1$, so in this case we choose $f_c=f_{ISCO}$. For simplicity we set the parameters of the minispike the same for different central IMBH. In the three figures we can see that for large $\alpha$ and small mass of IMBH the time difference is significant. In three cases for $\alpha=7/3$ the DM effect should be distinguished by LISA, i.e, for Fig.\ref{fig01}(c) the time difference is larger than 2 hours for a 4 year observation. However, when $\alpha$ is small, the time difference is insignificant, and an accurate estimation of the DM minispike effect relies on the accurate phase difference.

We define the phase difference $\triangle\tilde{\Phi}$ by
\begin{equation}
\Delta \tilde{\Phi}(f)=\tilde{\Phi}(f)-\tilde{\Phi}_0(f),\label{e54}
\end{equation}
which is the phase difference with and without DM minispike. From Eq.(\ref{e41}) we can see that $\Delta\tilde{\Phi}$ actually represents $\Delta\Psi$.

\begin{figure}[!ht]
\centering
\includegraphics[width=9.5cm]{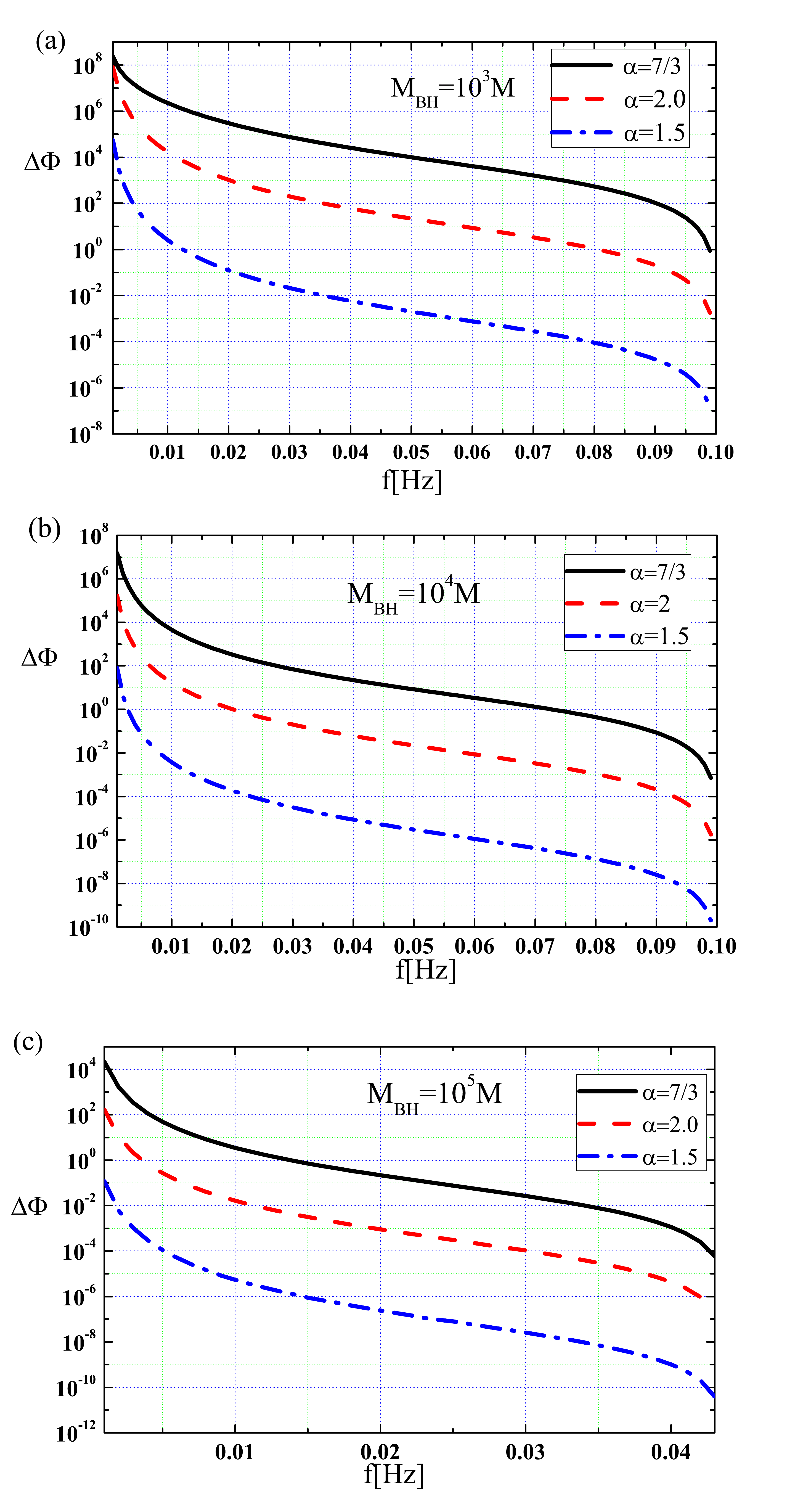}
\caption{The phase difference with different central IMBH. The horizontal axis is the initial frequency $f$ and the vertical axis is the phase difference. We take $\mu_0=10M_{\odot}$, $\rho_{\rm sp}=226M_\odot/pc^3$, $r_{\rm sp}=0.54pc$. The solid line $\alpha=7/3$, the dashed line $\alpha=2.0$, the dash-dot line $\alpha=1.5$. In figure(a) $M_{\rm BH}=10^3M_{\odot}$ and $f_c=0.1Hz$, in figure(b) $M_{\rm BH}=10^4M_{\odot}$ and $f_c=0.1Hz$, in figure(c) $M_{\rm BH}=10^5M_{\odot}$ and $f_c=f_{ISCO}$.} \label{fig02}
\end{figure}

Figure\ref{fig02} shows the phase difference with different masses of central IMBH in  frequency domain. Figure\ref{fig02}(a)(b) the frequency varies from $f=0.001$ to $f_c=0.1$ and Fig.\ref{fig02}(c) from $f=0.001$ to $f=f_{ISCO}$. As in Fig.\ref{fig01} the phase difference is significant with large $\alpha$ and small masses of central IMBH. It can be seen that in the frequency range depicted for $\alpha>1.5$ the DM effect can be distinguished by LISA, for example in Fig.\ref{fig02}(c) the minimum of the phase difference is $\Delta\tilde{\Phi}\sim 0.1$ with initial frequency $f=0.001$. However, if we consider the actual observation, the frequency range cannot be covered totally by a specific observation as the lifetime of LISA is about 4-5 years. One can verify that with $M_{\rm BH}=10^3M_{\odot}$ and $f_c=0.1Hz$, the initial frequency corresponding to $t_0=4year$ is about $f\sim 0.01$, and from Fig.\ref{fig01}(a) the phase difference is also a significant value for all three $\alpha$. For $M_{\rm BH}=10^4M_{\odot}$, $f\sim 0.006$ and for $M_{\rm BH}=10^5M_{\odot}$ $f\sim 0.003$. In the $M_{\rm BH}=10^5$ case the phase difference is insignificant for $\alpha=1.5$.

At last, we extract the phase difference by accretion specifically. If the accretion effect is not considered, the phase is
\begin{eqnarray}
&&\tilde{\Phi}_1=\frac{10}{3}\left(\frac{8\pi GM_{c0}}{c^3}\right)^{-5/3}\left[-f\int_{f_{c}}^f df'f'^{-11/3}L'^{-1}(f')\right.\nonumber\\
&&\left.+\int_{f_{c}}^f df'f'^{-8/3/3}L'^{-1}(f')\right],\label{e55}\\
&&L'(f)=1+4c_\epsilon\tilde{\delta}^{(11-2\alpha)/[2(3-\alpha)]}.\label{e56}
\end{eqnarray}
We define the phase difference caused by accretion $\delta \Phi$ by
\begin{eqnarray}
\delta\tilde{\Phi}(f)=\tilde{\Phi}(f)-\tilde{\Phi}_1(f).\label{e56}
\end{eqnarray}
\begin{figure}[!h]
\centering
\includegraphics[width=9.5cm]{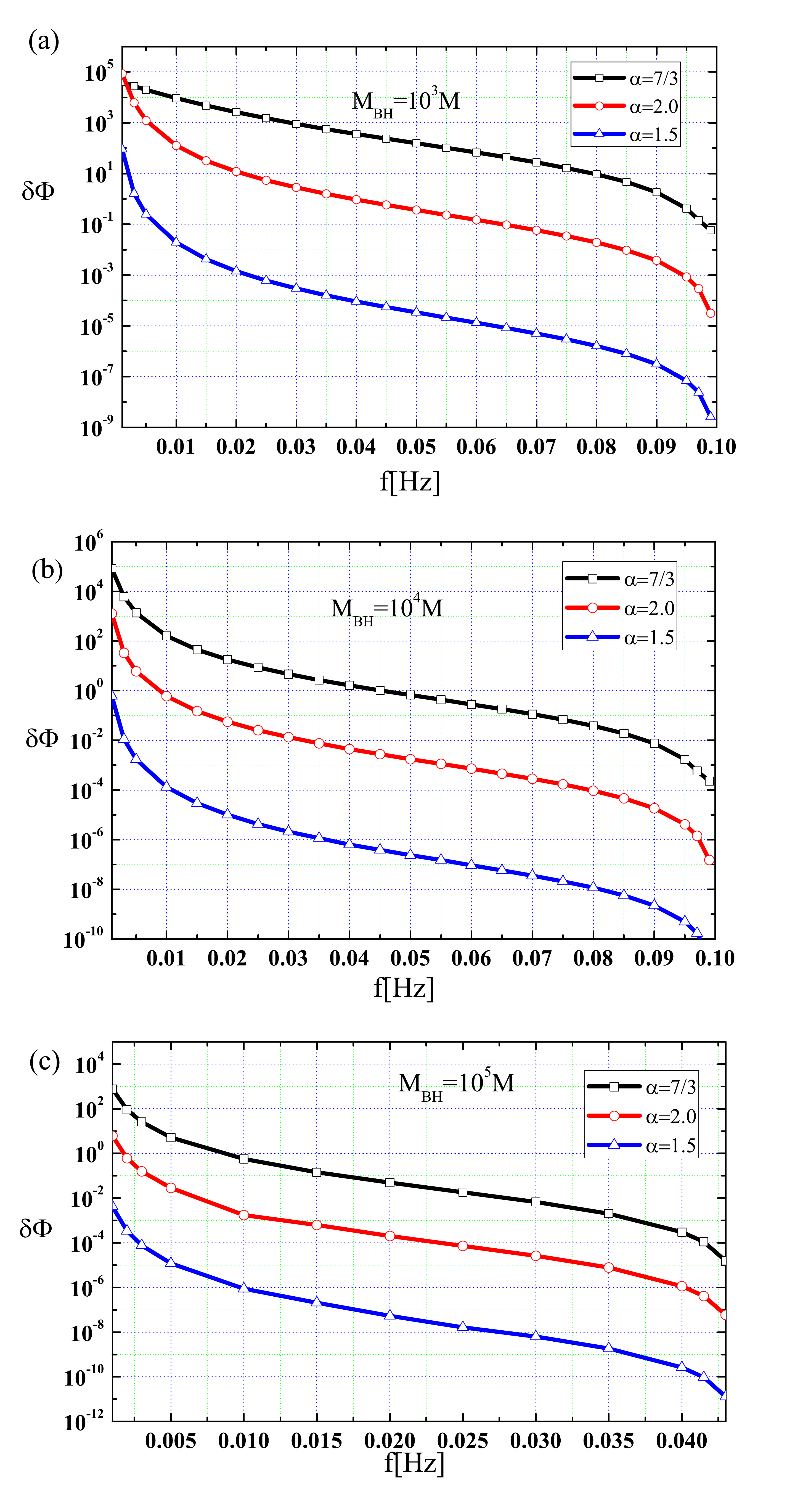}
\caption{The phase difference by accretion with different central IMBH. We take $\mu_0=10M_{\odot}$, $\rho_{\rm sp}=226M_\odot/{\rm pc}^3$, $r_{\rm sp}=0.54 $pc.In figure(a) $M_{BH}=10^3M_\odot$ and  $f_c=0.1Hz$, in figure(b) $M_{BH}=10^4M_\odot$ and $f_c=0.1Hz$, in figure(c) $M_{BH}=10^5M_\odot$ and $f_c=f_{ISCO}$. }\label{fig03}
\end{figure}

The phase difference $\delta\tilde{\Phi}$ are shown in Fig.\ref{fig03}. Compare Fig.\ref{fig02} and Fig.\ref{fig03} we can find that the phase shift caused by accretion $\delta\tilde{\Phi}$ is subordinate in the whole phase shift $\Delta\tilde{\Phi}$, for example, in Fig.\ref{fig02}(a) for $\alpha=7/3$, $\Delta\tilde{\Phi}\sim 10^4$ at $f=0.05Hz$ while $\delta\tilde{\Phi}\sim 10^2$ in Fig.\ref{fig03}(a). The ratio $\delta\tilde{\Phi}/\Delta\tilde{\Phi}$ is about $10^{-2}\sim 10^{-3}$.  In fact, one can verify that with $M_{\rm BH}=10^3M_\odot$ and $\alpha=7/3$, from $f=10^{-3}Hz$ to $0.1Hz$ the ratio of the accreted mass to the initial mass of the small BH $\Delta\mu/\mu_0\sim 3\times 10^{-3}$, so the accretion influence to the orbit is very weak. However, the accumulated phase shift $\delta\tilde{\Phi}$ is significant when $\Delta\Phi$ is large, as in the case for large $\alpha$ and small $M_{\rm BH}$. In \cite{I12} it was shown that the effect of friction overwhelms that of gravitational pull and the order of phase difference due to gravitational pull is about 5 orders lower than that of friction. In this paper we can see that when the combined effect including accretion is considered, the effect of dynamical friction is also dominant, but the accretion effect is much more significant than gravitational pull.

\begin{figure}[!h]
\centering
\includegraphics[width=9.0cm]{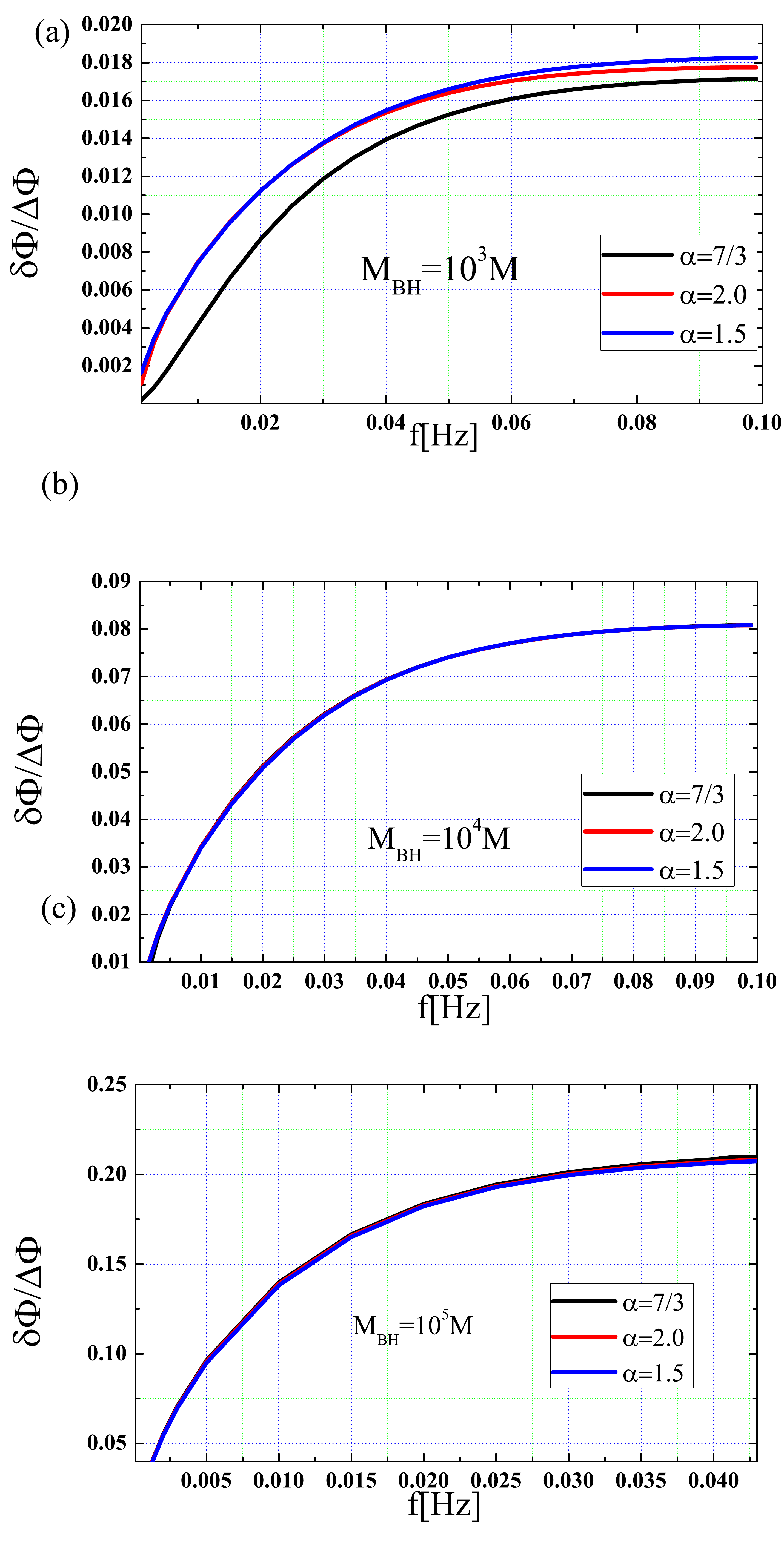}
\caption{The proportion of phase difference caused by accretion to total dephasing. Curves for different $\alpha$ overlap in panels (b, c). We take $\mu_0=10M_{\odot}$, $\rho_{\rm sp}=226M_\odot/{\rm pc}^3$, $r_{\rm sp}=0.54{\rm pc}$. In figure(a) $M_{BH}=10^3M_\odot$ and  $f_c=0.1Hz$, in figure(b) $M_{BH}=10^4M_\odot$ and $f_c=0.1Hz$, in figure(c) $M_{BH}=10^5M_\odot$ and $f_c=f_{ISCO}$ }\label{fig04}
\end{figure}

We present the proportion of accretion effect comparing to the total one in Fig. \ref{fig04}. We find that the accretion becomes more important while the mass of IMBH is larger. For the $M_{\rm BH}= 10^5 m_\odot$ case, the accretion effect accounts for a considerable proportion, as large as 20\%. Interestingly, the proportion looks insensitive with the density profile $\alpha$. In the panels (b, c), the curves represented different $\alpha$ overlap together. 

\begin{figure}[!h]
\centering
\includegraphics[width=9.5cm]{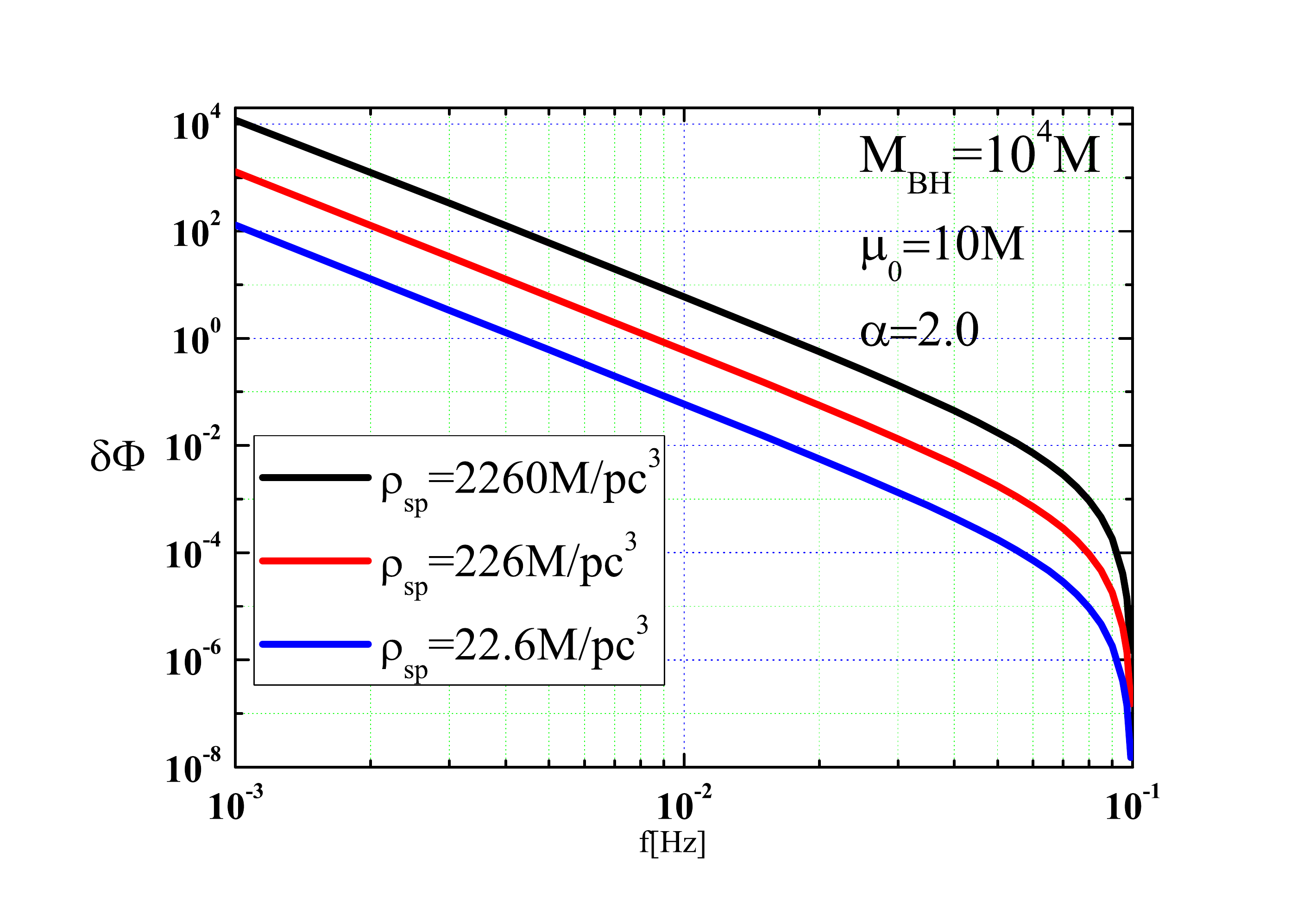}
\includegraphics[width=9.5cm]{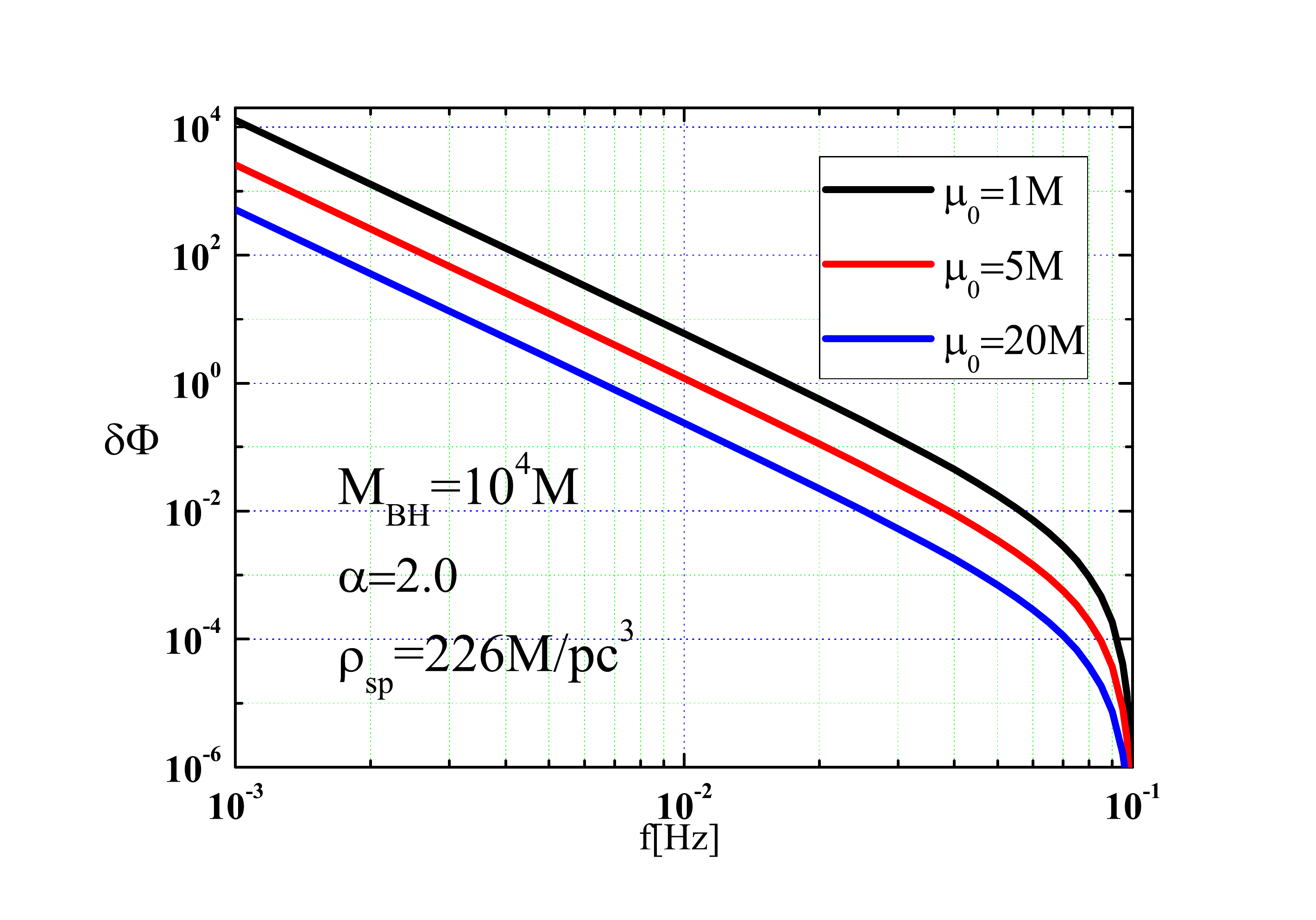}
\caption{Top panel: the phase difference caused by accretion with different density $\rho_{\rm sp}$. We set $M_{\rm BH}=10^4M, ~\mu_0=10M, ~r_{\rm sp}=0.5$ pc  and $\alpha=2.0$.   Bottom panel: the phase difference caused by accretion with different masses $\mu_0$ of the small black holes . We set $M_{\rm BH}=10^4M$,$r_{\rm sp}=0.5$ pc,  $\rho_{\rm sp}=226M/pc^3$ and $\alpha=2.0$. }\label{fig56}
\end{figure}

It is easily to deduce that the larger DM density $\rho_{\rm sp}$, the larger influence of minispike will be. This point is demonstrated in the top panel of Fig. \ref{fig56}. The depahsing is proportional to the value of $\rho_{\rm sp}$. We also reveal the accretion effect on dephasing by changing the mass of small BH. In the bottom panel of Fig. \ref{fig56}, we find that for the  smaller black hole, the phase difference is larger. This is reasonable, because the gravitational radiation is smaller if stellar BH is lighter, then the small black hole will experience more orbital cycles and longer evolution time. As a result, the influence of DM will be more obvious.

\section{\label{s4}Summary and conclusions}
In this paper we study the GW of an intermediate-mass-ratio inspirals with a central IMBH in a DM minispike analytically and numerically. As the accretion in the DM minispike is inevitable, we calculated the GW waveform of this system with a consideration of the combined effect of gravitational pull, dynamical friction and accretion. Employing the power law model of the minispike in\cite{I12}, we derived the GW waveform equations.

 With the numerical method we proposed a visualized description of the time difference in Fig.\ref{fig01} for a 4 year observation. Assuming the same DM minispike parameters $\rho_{\rm sp}$ and $r_{\rm sp}$ for convenience, we compare the time difference with different masses of central IMBH and different power law $\alpha$ of the DM minispike. We find that for large $\alpha$ and smaller mass of the central BH, the DM effect makes the time difference significant and is distinguishable by LISA. In the case of central IMBH with mass of $10^5M_\odot$, the advanced time of merger can also be detectable by LISA for large $\alpha$.However, when $\alpha$ is small and the mass of central IMBH is large, the time difference is insignificant. As the sensitivity of the LISA to the GW phase, we  have to turn to the phase difference to see the effect of DM minispike.

 We also proposed the phase difference with the effect including accretion for different central IMBH in Fig.\ref{fig02}. With the frequency range $0.001Hz\sim 0.1Hz$, the phase difference shows that for $\alpha>1.5$ the effect of DM minispike can be distinguished considering the accuracy of LISA. However,  for a feasible observation of 4-5 years, the frequency range can not be totally covered, and as a result for the small $\alpha$ and large $M_{\rm BH}$ the phase difference is insignificant. On the other hand, when the central IMBH has a larger mass, the density and the total mass of the minispike could not be the same as the smaller mass case as we assumed and should be larger.  The phase difference should be more significant than the results depicted as well. The relation between the DM density and the mass of central black hole is diverse for different DM models, which is not concern in this paper. But we  can still conclude that with the same DM density the larger mass of the central IMBH can weaken the effect of DM minispike.  As a result, in the detection of DM with GW  we should consider the comprehensive effect of the mass of the central IMBH.

We extract the phase difference caused by accretion specifically  and compare it with that by friction and gravitational pull.  Compared with dynamical friction the accretion drag is small. The accreted mass is also a small quantity compared  with the initial mass of the small BH and the influence of accretion to the orbit is very weak, at least in the case of our consideration.  The numerical results Fig.\ref{fig03} show that in the whole phase difference $\Delta\tilde{\Phi}$ the accretion contribution is inconspicuous.  However, the numerical results  shows that the phase shift  by accretion $\delta\tilde{\Phi}$ can be a large quantity when $\Delta\Phi$ is large. Figure \ref{fig04} shows that the proportion of accretion can be as large as 20\%. We can conclude that in the effect of DM minispike the dynamical friction is dominant, and the order of magnitude estimation shows that the accretion effect is much more significant than that of gravitational pull. Obviously the accretion effect can not be ignored in the detection of GW and the phase shift is so large that we have to use the waveform including the accretion effect as a template.  Also, we reveal that the mass of the stellar BH is smaller, the accretion effect is larger.

In \cite{I11} it was concluded that the environmental effects can be negligible for GW detection of LISA. In this paper we find again that in accordance of the huge number of orbital cycles which the binary experienced in LISA frequency band, a very tiny effect could have a large impact on the detection of GW, as pointed in \cite{I100}. Only the order of magnitude estimates may be not sufficient to determine whether an effect can be detectable.

In \cite{I12}, it was indicated that the DM parameters can be extracted very accurately from the GW waveform using the filtering technique and Fisher matrix analysis. In Sec.\ref{s3C} we can see that the accretion effect makes the equations of GW waveform complicated. When we take into account of the accretion effect, the chirp mass $M_c$ varies with time as the accretion of the small BH. Whether the accretion makes the estimation of the DM parameters complicated and how accurately we can extract the parameters deserves more investigations.

The GW observation is a powerful tool to detect environments around BH and other stellar objects. In this paper we concentrate on the nonannihilating DM particles. If we consider other natures of DM, the situation becomes complicated. As we have to consider the interactions beyond gravitation, the accretion and dynamical friction may change their forms from that in this paper, which is an open issue. On the other hand, the observations of other environments such as accretion disks, magnetic fields by GW detection is not studied systematically, and requires future investigation.

\begin{acknowledgments}
This work is supported by National Natural Science Foundation of China, No. 11773059, No. U1431120, No. 11690023; and by Key Research Program of Frontier Sciences, CAS, No. QYZDB-SSW-SYS016. WH is also supported by the Youth Innovation Promotion Association of CAS.

\end{acknowledgments}

\end{document}